\documentstyle[preprint,aps,prbbib]{revtex}                                                 

\begin{document}     

\begin{center}
{\Large\bf Velocity, Acceleration and Cosmic Distances in Cosmological Special
Relativity}

\end{center}

\begin{center}
{Moshe Carmeli}
\end{center}

\begin{center}
{Department of Physics, Ben Gurion University, Beer Sheva 84105, 
Israel}
\end{center}

\begin{center}
{Email: carmelim@bgumail.bgu.ac.il}
\end{center}

\begin{abstract}
In this paper we present the fundamentals of the cosmological special 
relativity (CSR) [1-3] by discussing the {\it dynamical} concepts of 
{\it velocity},
{\it acceleration} and {\it cosmic distances} in spacevelocity. These concepts 
occur in CSR just as those of mass, linear momentum and energy appear in 
Einstein's special relativity (ESR) of spacetime (see Chapter 7 of Ref. 1). 
\end{abstract}
\newpage
\section{Preliminaries}
The most important result of Einstein's special relativity is probably the
relationship between mass and energy (see Chapter 7 of Ref. 1). How it 
happened that the
mass became so critical in this theory? The answer is very simple. The theory
involves the square of the speed of light, $c^2$. What physical quantity 
incorporates the square of velocity?
It is the energy, $mv^2/2$. Hence it is the mass that goes with $v^2$ and 
$c^2$. Thus $m$ is taken as an invariant under the Lorentz transformation. 
This becomes the rest mass $m_0$. The inertial mass $m$ follows to depend on 
the velocity.

What is the comparable physical quantity in cosmological special relativity?
Certainly not the mass. Here we have $\tau^2$, the square of the Hubble time
constant. What physical quantity goes with the square of time? It is the 
acceleration {\bf a} because ${\bf a}t^2/2$ describes the distance a particle
makes at time $t$ when it is subject to acceleration {\bf a}. Hence {\bf a}
should be taken as the invariant quantity under the cosmological 
transformation. But then the acceleration depends on the cosmic time just as
the mass depends on the velocity.

In this paper these concepts are explored.
\section{Velocity and acceleration four-vectors}

We start our four-dimensional spacevelocity analysis by defining the velocity
and acceleration (see Section 5.5 of Ref. 1 for the special relativistic 
treatment in spacetime).

The {\it velocity} four-vector of a particle in spacevelocity is defined as the
dimensionless quantity
$$u^\mu=\frac{dx^\mu}{ds},\eqno(1)$$ 
where $\mu=0,1,2,3,$ $x^\mu=(x^0,x^1,x^2,x^3)=(\tau v,x,y,z)$, and $\tau$ is
Hubble's time in the limit of zero gravity, a universal constant whose value
is $12.5\times 10^9$ years [4,5]. In flat spacevelocity one has for the line element,
$$\tau^2dv^2-\left(dx^2+dy^2+dz^2\right)=ds^2,\eqno(2)$$
thus
$$\tau^2\left(\frac{dv}{ds}\right)^2\left(1-\frac{dx^2+dy^2+dz^2}{\tau^2dv^2}
\right)=1.\eqno(3)$$
This gives
$$\tau^2\left(\frac{dv}{ds}\right)^2\left(1-\frac{t^2}{\tau^2}\right)=1,
\eqno(4)$$
and therefore
$$\frac{dv}{ds}=\frac{1}{\tau\sqrt{1-
\displaystyle\normalsize\frac{t^2}{\tau^2}}}.\eqno(5)$$
The {\it velocity} four-vector in spacevelocity can thus be expressed as 
$$u^\mu=\frac{dx^\mu}{ds}=\frac{dx^\mu}{dv}\frac{dv}{ds}=
\frac{1}{\tau\sqrt{1-
\displaystyle\normalsize\frac{t^2}{\tau^2}}}\frac{dx^\mu}{dv}.\eqno(6)$$

The velocitylike component of $u^\mu$ is therefore given by
$$u^0=\gamma,\eqno(7)$$
whereas its spatial components
$$\mbox{\bf u}=u^k=\left(u^1,u^2,u^3\right) \eqno(8)$$
are given by
$$u^k=\frac{\gamma}{\tau}\frac{dx^k}{dv},\eqno(9)$$
where
$$\gamma=\frac{1}{\sqrt{1-
\displaystyle\normalsize\frac{t^2}{\tau^2}}}.\eqno(10)$$

It will be noted that, by Eq. (1),
$$u_\alpha u^\alpha=1,\eqno(11)$$
namely, the length of $u^\mu$ is unity.

The {\it acceleration} four-vector of a particle in spacevelocity is defined 
by
$$\frac{du^\mu}{ds}=\frac{d^2x^\mu}{ds^2}.\eqno(12)$$
By differentiating Eq. (11) we find that the acceleration four-vector 
satisfies the orthogonality condition
$$u_\alpha\frac{du^\alpha}{ds}=0.\eqno(13)$$

The components of the acceleration four-vector of a particle in spacevelocity
are, by Eqs. (5) and (7)--(9), then given by
$$\frac{du^0}{ds}=\frac{du^0}{dv}\frac{dv}{ds}=
\frac{\gamma}{\tau}\frac{d\gamma}{dv},\eqno(14)$$
\vspace{0mm}
$$\frac{du^k}{ds}=\frac{du^k}{dv}\frac{dv}{ds}=\frac{\gamma}{\tau^2}
\frac{d}{dv}\left(\gamma\frac{dx^k}{dv}\right),\eqno(15)$$
where $\gamma$ is given by Eq. (10).
\section{Acceleration and cosmic distances}
Multiplying Eq. (4) by $\mbox{\bf a}_0^2\tau^4$, where $\mbox{\bf a}_0$ is
the ordinary three-vector acceleration as measured in the cosmic frame of 
reference (see Section 2.5 of Ref. 1) at cosmic time $t=0$ (i.e. now). 
We  obtain
$$\mbox{\bf a}_0^2\tau^6\left(\frac{dv}{ds}\right)^2\left(1-\frac{t^2}{\tau^2}
\right)=\mbox{\bf a}_0^2\tau^4.\eqno(16)$$
Using now Eq. (5) in Eq. (16) the latter then yields
$$\left(1-\frac{t^2}{\tau^2}\right)^{-1}\left(\mbox{\bf a}_0^2
\tau^4-\mbox{\bf a}_0^2t^2\tau^2\right)=\mbox{\bf a}_0^2\tau^4.\eqno(17)$$

We now define the acceleration {\bf a} at an arbitrary cosmic time $t$ by
$$\mbox{\bf a}=\frac{\mbox{\bf a}_0}{\sqrt{1-
\displaystyle\normalsize\frac{t^2}{\tau^2}}},\eqno(18)$$
then Eq. (17) will have the form
$$\mbox{\bf a}^2\tau^4-\mbox{\bf a}^2t^2\tau^2=\mbox{\bf a}_0^2\tau^4,
\eqno(19)$$
or   
$$\mbox{\bf a}^2\tau^4-\tau^2\mbox{\bf v}^2=\mbox{\bf a}_0^2\tau^4,
\eqno(20)$$
where $\mbox{\bf v}=\mbox{\bf a}t$ is the ordinary three-dimensional velocity.
Equation (20) is the analog to 
$$m^2c^4-c^2\mbox{\bf p}^2=m_0^2c^4,\eqno(21)$$
in ESR, where $\mbox{\bf p}=m\mbox{\bf v}$ is the linear momentum (see Section
7.2 of  Ref. 1). Thus {\bf a} reduces to $\mbox{\bf a}_0$ when the cosmic time $t=0$ 
(i.e. at present).

The comparable to Eq. (18) in ESR is, of course,
$$m=\frac{m_0}{\sqrt{1-
\displaystyle\normalsize\frac{v^2}{c^2}}},\eqno(22)$$
where $m$ and $m_0$ are the inertial mass and the rest mass of the particle,
with $m$ reduces to $m_0$ at $v=0$. If we multiply Eq. (22) by $c^2$ and 
expand both sides in $v/c$, we obtain
$$mc^2=m_0c^2+\frac{m_0}{2}v^2+\cdots .\eqno(23)$$
Doing the same with Eq. (18) but multiplication by $\tau^2$, and expanding 
in $t/\tau$, we obtain
$$\mbox{\bf a}\tau^2=\mbox{\bf a}_0\tau^2+\frac{\mbox{\bf a}_0}{2}t^2+\cdots .
\eqno(24)$$
Equation (24) in CSR is of course the analog to Eq. (23) in ESR.
\section{Energy in ESR versus cosmic distance in CSR}
While Eq. (23) yields
$$E=E_0+\frac{m_0}{2}v^2+\cdots,\eqno(25)$$
Eq. (24) gives
$$\mbox{\bf S}=\mbox{\bf S}_0+\frac{\mbox{\bf a}_0}{2}t^2+\cdots.\eqno(26)$$
In the above equations $E$ is, of course, the energy of the particle, whereas
{\bf S} is the cosmic distance. $E_0=m_0c^2$ is the rest energy of the 
particle whereas $m_0v^2/2$ is the Newtonian kinetic energy. What about the
terms in Eq. (26)? The term $\mbox{\bf a}_0t^2/2$ is, of course, the 
Newtonian distance the particle makes due to the acceleration $\mbox{\bf a}_0$,
the term $\mbox{\bf S}_0=\mbox{\bf a}_0\tau^2$ is unique to CSR, and it might
be called the {\it intrinsic} cosmic distance of the particle. 

Equation (20) can now be written as
$$\mbox{\bf S}^2-\tau^2\mbox{\bf v}^2=\mbox{\bf S}_0^2,\eqno(27)$$
in complete analogy to
$$E^2-c^2\mbox{\bf p}^2=E_0^2 \eqno(28)$$
in ESR.
\section{Cosmic distance-velocity four-vector}
We now define the cosmic distance-velocity four-vector. It  is the analogous 
to the energy-momentum four-vector in ESR (see Section 7.4 of Ref. 1). It is 
defined by
$$\mbox{\bf v}^\mu=\mbox{\bf a}_0\tau^2u^\mu,\eqno(29)$$
where $u^\mu$ has been defined in Section 2.
We have 
$$\mbox{\bf v}^0=\mbox{\bf a}_0\tau^2u^0,\eqno(30a)$$
$$\mbox{\bf v}^k=\mbox{\bf a}_0\tau^2u^k,\eqno(30b)$$
where $u^0$ and $u^k$ are given by Eqs. (7)--(10), with
$$\mbox{\bf v}_0=\mbox{\bf v}^0,\hspace{5mm} \mbox{\bf v}_k=-\mbox{\bf v}^k.
\eqno(31)$$

Accordingly we have 
$$\mbox{\bf v}^2=\mbox{\bf v}_0\cdot\mbox{\bf v}^0+\mbox{\bf v}_k\cdot\mbox{\bf v}^k=
\mbox{\bf a}_0^2\tau^4\left(u_0u^0+u_ku^k\right)$$
$$=\mbox{\bf a}_0^2\tau^4
u_\alpha u^\alpha=\mbox{\bf a}_0^2\tau^4=\mbox{\bf S}_0^2.\eqno(32)$$
But, using Eqs. (7), (9) and (18),
$$\mbox{\bf v}^0=\mbox{\bf a}_0\tau^2\gamma=\mbox{\bf a}\tau^2=\mbox{\bf S},
\eqno(33a)$$
$$\mbox{\bf v}^k=\mbox{\bf a}_0\tau\gamma\frac{dx^k}{dv}=\mbox{\bf a}\tau
\frac{dx^k}{dv}=\mbox{\bf a}\tau\frac{dx^k}{dt}\frac{dt}{dv}=
\tau\mbox{\bf v},\eqno(33b)$$
where {\bf v} is the three-dimensional velocity. Hence
$$\mbox{\bf v}_\alpha\cdot\mbox{\bf v}^\alpha=\mbox{\bf v}_0\cdot\mbox{\bf v}^0
+\mbox{\bf v}_k\cdot\mbox{\bf v}^k=\mbox{\bf a}_0^2\tau^4\gamma^2-\mbox{\bf a}_0^2
\tau^2\gamma^2\left(\frac{dx^k}{dv}\right)^2$$
$$=\mbox{\bf a}^2\tau^4-\tau^2
\mbox{\bf v}^2=\mbox{\bf S}^2-\tau^2\mbox{\bf v}^2,\eqno(34)$$
and accordingly, using (32),
$$\mbox{\bf S}^2-\tau^2\mbox{\bf v}^2=\mbox{\bf S}_0^2,\eqno(35)$$
which is exactly Eq. (20) with $\mbox{\bf S}=\mbox{\bf a}\tau^2$ and 
$\mbox{\bf S}_0=\mbox{\bf a}_0\tau^2$. 
The above analysis also shows that Eq. (35) is covariant under spacevelocity 
cosmological transformation.

Equation (35) is the analog to 
$$E^2-c^2\mbox{\bf p}^2=E_0^2 \eqno(36)$$
in ESR, where $E$ and $E_0$ are the energy and rest energy, respectively, 
{\bf S} and $\mbox{\bf S}_0$ are the cosmic distances at cosmic time $t$ and present
time ($t=0$), respectively. 

Finally, in ESR when the rest mass is zero (like the photon), one then has
$$E=cp,\eqno(37)$$
which is valid at the light cone (see Chapter 6 of Ref. 1). In the case of CSR one also
has, when $\mbox{\bf S}_0=0$,
$$\mbox{\bf S}=\tau\mbox{\bf v},\eqno(38)$$
now valid at the galaxy cone (see Section 2.14 of Ref. 1).
\section{Conclusions}
In this paper it has been shown that the comparable quantity to the mass in
Einstein's special relativity is the ordinary acceleration three-vector in
cosmological special relativity. They both have similar behavior, one with 
respect to $v/c$ and the other with $t/\tau$:
$$m=\frac{m_0}{\sqrt{1-\displaystyle\normalsize\frac{v^2}{c^2}}},\eqno(39)$$
and
$$\mbox{\bf a}=\frac{\mbox{\bf a}_0}{\sqrt{1-
\displaystyle\normalsize\frac{t^2}{\tau^2}}}.\eqno(40)$$

Furthermore, the role of the energy in Einstein's theory is being taken over 
by the cosmic distance,
$$E=mc^2,\eqno(41)$$
and
$$\mbox{\bf S}=\mbox{\bf a}\tau^2.\eqno(42)$$

Finally, the analog of the energy formula 
$$E^2-c^2\mbox{\bf p}^2=E_0^2\eqno(43)$$
in ordinary special relativity is
$$\mbox{\bf S}^2-\tau^2\mbox{\bf v}^2=\mbox{\bf S}_0^2\eqno(44)$$
in cosmological special relativity.

\end{document}